\documentclass{aa}

\usepackage{graphicx,astron,epsfig}

\def\be {\begin{equation}}
\def\en {\end{equation}}
\def\bea {\begin{eqnarray}}
\def\ena {\end{eqnarray}}
\def\bi {\begin{itemize}}
\def\ei {\end{itemize}}
\def\eg {{\it e.g. }}
\def\ie {{\it i.e. }}

\begin{document}

\thesaurus{02(12.03.1;03.13.2)}

\title{MAPCUMBA : a fast iterative multi-grid map-making algorithm for
CMB experiments} 

\author{O. Dor{\'e}\inst{1} \and R. Teyssier\inst{1,2} \and
F.R. Bouchet\inst{1} \and D. Vibert\inst{1} \and S. Prunet\inst{3} }

\institute{Institut d'Astrophysique de Paris, 98bis, Boulevard Arago,
75014 Paris, FRANCE \and Service d'Astrophysique, DAPNIA, Centre
d'{\'E}tudes de Saclay, 91191 Gif-sur-Yvette, FRANCE \and CITA,
McLennan Labs, 60 St George Street, Toronto, ON M5S 3H8, CANADA}

\offprints{O.~Dor{\'e}}
\mail{dore@iap.fr}
\authorrunning{Dor{\'e}, Teyssier, Bouchet, Vibert \& Prunet}
\titlerunning{MAPCUMBA: a fast multi-grid CMB map-making code}

\maketitle
\begin{abstract}

The data analysis of current Cosmic Microwave Background
(CMB) experiments like BOOMERanG or MAXIMA poses severe challenges
which already stretch the limits of current (super-) computer
capabilities, if brute force methods are used. In this paper we present
a practical solution to the optimal map making problem which can be
used directly for next generation CMB experiments like ARCHEOPS and
TopHat, and can probably be extended relatively easily to the full
PLANCK case. This solution is based on an iterative multi-grid Jacobi
algorithm which is both fast and memory sparing. Indeed, if there are
$\mathcal{N}_{tod}$ data points along the one dimensional timeline to analyse,
the number of operations is of $\mathcal{O} (\mathcal{N}_{tod} \ln
\mathcal{N}_{tod})$ and the memory requirement is $\mathcal{O}
(\mathcal{N}_{tod})$. Timing and accuracy issues have been analysed on
simulated ARCHEOPS and TopHat data, and we discuss as well the issue
of the joint evaluation of the signal and noise statistical
properties.

\end{abstract}

\section{Introduction}

As cosmology enters the era of ``precision'', it enters simultaneously
the era of massive data sets. This has in turn showed the need for new
data processing algorithm. Present and future CMB experiments in
particular face some new and interesting challenges \cite{BoCr99}. If
we accept the now classical point of view of a four steps data
analysis pipeline : i/ from \emph{time-ordered data} (TOD) to maps of
the sky at a given frequency, ii/ from frequency maps to (among
others) a temperature map, iii) from a temperature map to its power
spectrum $C_{\ell}$, iv/ from the power spectrum to cosmological
parameters and characteristics of the
primordial spectrum of fluctuation, the ultimate quantities to be
measured in a given model. The work we are presenting focus on the
first of these issues, namely the map-making step.

Up to the most recent experiments, maps could be made by a brute force
approach amounting to solve directly a large linear problem by direct
matrix inversion. Nevertheless the size of the problem, and the
required computing power, grows as a power law of the data set size,
and the limits of this kind of method have now been reached
\cite{Bo00}. Whereas the most efficient development in this massive
computing approach, \ie the MADCAP package \cite{Bo99} has been
applied to the recent BOOMERanG and MAXIMA experiments
\cite{dBe00,Ha00} some faster and less consuming solutions
based on iterative inversion algorithms have now been developed and
applied too \cite{PrNe00}. This is in the same spirit as
\cite{WrHi96a}, and we definitely follow this latter approach.

Design goals are to use memory sparingly by handling only columns or
rows instead of full matrixes and to increase speed by minimising the
number of iterations required to reach the sought convergence of the
iterative scheme. We fulfill these goals by an iterative multi-grid
Jacobi algorithm. As recalled below, an optimal method involves using
the noise power spectrum. We have thus investigated the 
possibility of a joint noise (and signal) evaluation using this algorithm.

Section \ref{method} presents in detail the method and its
implementation, while section \ref{application} demonstrates its
capabilities by using simulations of two on-going balloon CMB
experiments,
ARCHEOPS\footnote{\texttt{http://www-crtbt.polycnrs-gre.fr/archeops/}}\cite{Be00}
and TopHat\footnote{\texttt{http://topweb.gsfc.nasa.gov/}}. The results are
discussed in section \ref{discussion}, together with the problem of
the evaluation of the noise properties, as well as possible extensions.

\section{The method} \label{method}

\subsection{Optimal map making}

The relation between the sky map we seek and the observed data stream
may be cast as a linear algebra system \cite{WrHi96a,Te97}. Let $_t$
and $_p$ indices denote quantities in the temporal and spatial
domains, and group as a data vector, $d_{t}$, and a noise vector the
temporal stream of collected data and the detector noise stream, both of
dimension $\mathcal{N}_{tod}$. We then have 
\be
d_{t} = A_{tp}x_{p} + n_{t} ,
\en
where $A_{tp}x_{p}$ is the signal vector given by the observation of
the unknown pixelised sky map, $x_{p}$, which has been arranged as a
vector of dimension $\mathcal{N}_{pix}$. The $\mathcal{N}_{tod} \times 
\mathcal{N}_{pix}$ ``observation'' matrix $A$ therefore encompasses the
scanning strategy and the beam pattern of the detector. 

In the following, we restrict to the case when the beam pattern is
symmetrical. We can therefore take $x_{p}$ to be a map of the sky which
has already been convolved with the beam pattern, and $A$ only
describes how this sky is being scanned. For the total power measurement
(\ie non-differential measurement) we are interested in here, the
observation matrix $A$ then has a single non-zero element per row,
which can be set to one if $d$ and $x$ are expressed in the same
units. The model of the measurement is then quite transparent: each
temporal datum is the sum of the pixel value to which the detector is
pointing plus the detector noise.

The map-making step then amounts to best solve for $x$ given $d$ (and some
properties of the noise). We shall seek a \emph{linear} estimator of $x_{p}$,
\be
\hat{x}_{p} = W_{pt} d_{t}\ .
\en
To motivate a particular choice of the $\mathcal{N}_{pix} \times 
\mathcal{N}_{tod}$ matrix $W$, a Bayesian approach is convenient.
Indeed we are seeking the optimal solution to this inversion problem which
maximises the probability of a deduced set of theory parameters (here the map
$x_p$) given our data ($d_t$) by maximising $\mathcal{P}(x|d)$. Bayes'
theorem simply states that
\be
\mathcal{P}(x|d) =
\frac{\mathcal{P}(d|x)\mathcal{P}(x)}{\mathcal{P}(d)} \: .
\en
If we \emph{do not assume any theoretical prior}, then
$x$ follows a uniform distribution as well as $d$. Therefore,
\bea
\mathcal{P}(x|d) & \propto &  \mathcal{P}(d|x) .
\ena
If we further assume that \emph{the noise is Gaussian}, we can write
\bea
\mathcal{P}(x|d) & \propto & exp(- n^T_tN^{-1}_{tt'}n_{t'}/2) \\
                 & \propto & exp(-(d-Ax)_t^T N^{-1}_{tt'} (d-Ax)_{t'}/2)\\
                 & \propto & exp(-\chi^2/2)\:
\ena
where $N_{tt'} = <nn^T>_{tt'}$ is the noise covariance matrix.
In this particular case, maximising $\mathcal{P}(x|y)$ amounts to
find the least square solution which was used to analyse the ``COBE''
data \cite{JaGu92},
\be \label{likelihood}
W = \left[A^T N^{-1} A\right]^{-1} A^T N^{-1}\: .
\label{eq:noprior}
\en
In this paper we will actually deal only with this
estimator. Nevertheless as a next iteration in the analysis process,
we could incorporate various theoretical priors by expliciting
$\mathcal{P}(x)$. For example, it is often assumed 
\emph{a Gaussian prior} for the theory, \ie $\mathcal{P}(x) \propto
exp(-x^T_pC^{-1}_{pp'}x_{p'}/2)$ where 
$C_{pp'}=\langle x_px^T_{p'} \rangle$ is the
signal covariance matrix. In that case the particular solution
turns out to be the Wiener filtering solution
\cite{ZaHo95,BoGi96,TeEf96,BoGi98}:
\be
W = \left[ C^{-1} + A^T N^{-1} A \right]^{-1} A^T N^{-1}\ .
\en
But this solution may always be obtained by a further (Wiener)
filtering of the COBE solution, and we do not consider it further. 

The prior-less solution demonstrates that as long as the (Gaussian)
instrumental noise is not white, a simple averaging (co-addition) of
all the data points corresponding to a given sky pixel is not
optimal. If the noise exhibits some temporal correlations, as induced
for instance by a low-frequency $1/f$ behavior of the noise spectrum
which prevails in most CMB experiments, one has to take into account
the full time correlation structure of the noise. Additionally this
expression demonstrates that even if the noise has a simple time
structure, the scanning strategy generically induces a non-trivial
correlation matrix $\left[ A^T N^{-1} A\right]^{-1}$ of the noise map.

Even if the problem is well posed formally, a quick look at the
orders of magnitude shows that the actual finding of a solution is non
trivial task. Indeed a brute force method aiming at inverting the full
matrix $\left[ A^T N^{-1} A\right]^{-1} \displaystyle$, an operation
scaling as $\mathcal{O}(\mathcal{N}_{pix}^3)$, is already hardly
tractable for present long duration balloon flights as MAXIMA,
BOOMERanG, ARCHEOPS or TopHat where $\mathcal{N}_{tod}\sim 10^6$ and
$\mathcal{N}_{pix} \sim 10^5$. It appears totally impractical for
PLANCK since for a single detector (amid $10$s)
$\mathcal{N}_{tod}\sim 10^9$ and $\mathcal{N}_{pix} \sim 10^7$! 

One possibility may be to take advantage of specific scanning
strategies, and actually solve the inverse of the convolution problem
as detailed in \cite{WaGo00}. This amounts to deduce the
map coefficients $a_{lm}$ in the spherical harmonic basis through a
rotation of a Fourier decomposition of the observed data.  The map
will then be a simple visualisation device, while the $a_{lm}$ would
be ready to use directly for a component separation (as in
\cite{BoGi96,TeEf96,BoGi98}) and the CMB power spectrum
estimate. While potentially very interesting, this approach will not
be generally applicable (at least efficiently), and we now turn to a
practical (general) solution of equation~(\ref{eq:noprior}) by
iterative means. 

\subsection{Practical implementation for large data sets}

We solve the map-making problem by adapting to our particular case the
general ``multi-grid method'' \cite{PrTe92}. Multi-grid methods are
commonly used to speed up the convergence of a traditional relaxation
method (in our case the Jacobi method, as in \cite{PrNe00}) defined at
resolution $\ell_{max}$ (see below). A set of recursively defined
coarser grids ($\ell < \ell_{max}$) are used as temporary
computational space, in order to increase the convergence rate of the
relaxation process.  To be fully profitable, this algorithm implies
for each resolution both a rebinning in space (resolution change) and
in time (resampling).

In this paper, we use the HEALPix pixelisation of the sphere
\cite{GoHi98}. In this scheme, the sphere is covered by 12 basic
quadrilaterals, further divided recursively into pixels of equal
area. The map resolution is labeled by $N_{side}$: the number of
pixels along the side of one basic quadrilateral. Hence, $N_{side}=1$
means that the sphere is covered by 12 large pixels only. The
number of pixels is given by $N_{pix}=12 N_{side}^2$. $N_{side}
= 256$ corresponds to a pixel size of $13.7$ arcmin. For practical
reasons, we need to define the {\it resolution} $k$ of a HEALPix
map as
\be
N_{side}=2^{k}
\en
The ``nested''pixel numbering scheme of HEALPix \cite{GoHi98} allows an
easy implementation of the coarsening ($k \rightarrow
k-1$) and refining ($k \rightarrow k+1$) operators that
we use intensively in our multigrid scheme.

Let us now get into the details of our implementation and discuss
successively the exact system we solve, the way we solve it and 
the actual steps of the multi-grid algorithm.

\subsubsection{Determining the working resolution $k_{max}$}
\label{lmax}

We aim at solving for the optimal map $\hat{x}_{k}$ at a given
spatial resolution $k$ using
\be \label{start}
A_{k}^T N^{-1} A_{k}\ \hat{x}_{k} = A_{k}^T N^{-1}\ d\: ,
\en
where $A_{k}$ is the ``observation'' operator (from spatial to
temporal domain) and $A_{k}^T$ is the ``projection'' operator
(from temporal to spatial domain). In a noise-free experiment, the
optimal map would be straightforwardly given by the co-added map
(introducing the ``co-addition'' operator $P_{k}$)
\be
\hat{x}_{k} = P_{k}\, d \equiv (A_{k}^T A_{k})^{-1} A_{k}^T\ d 
\en
The time line is given by $ d = A_{\infty} x_{\infty} + n$ where
$x_{\infty}$ is the sky map at ``infinite'' resolution
($k=+\infty$ in our notations). In order to check the accuracy
of this trivial noise-free map making, it is natural to compute the
residual {\it in the time domain} with $n=0$
\be
p_{k} = A_{k} \hat{x}_{k} - d
= A_{k}\hat{x}_{k} - A_{\infty} x_{\infty}
\en
which will be non-zero in practice, as soon as one works with
finite spatial resolution. We call this residual the {\it
pixelisation noise}. Since we assume here that the instrumental
beam is symmetric, the sky map is considered as the true sky
convolved by, say, a Gaussian beam of angular diameter $\Delta
\theta_B$. This introduces a low-pass spatial filter in the problem.
In other words, as the resolution increase, the pixelisation noise
should decrease towards zero. We have estimated that the order of
magnitude of the pixelisation noise can be approximated by
\be
\| p_{k} \| \simeq \| x_{\infty} \| \frac{ \Delta \theta _{k} }{ \Delta \theta _B  }
\en
The norms used in the above formula can be either the maximum over
the time line (a very strong constraint) or the variance over the
time line (a weaker constraint). Since the pixelisation noise is
strongly correlated with the sky signal, point sources or Galaxy
crossings are potential candidates for large and localised bursts
of pixelisation noise. The correct working resolution $k_{max}$
is set by requiring that the pixelisation noise remains low
compared to the actual instrumental noise, or equivalently
\be
\Delta \theta _{k_{max}} \le \frac{ \Delta \theta _B }{S/N }
\en
Most of the CMB experiments are noise dominated along the time
line, constraining the effective map resolution to be of the order
of the instrumental beam or even larger. Note however that the
pixelisation noise is strongly non Gaussian (point sources or
Galaxy crossings) and can be always considered as a potential
source of residual stripes in the final maps.

\subsubsection{The basic relaxation method: an approximate Jacobi solver }
\label{noiseiter}

Instead of solving for $\hat{x}$ we perform the change of
variable
\be
\hat{y} = \hat{x} - P\ d
\label{noisedef}
\en
and solve instead for $\hat{y}$ which obeys
\be \label{eqchgt}
P N^{-1} A\ \hat{y} = P N^{-1} (d - A P d)~~~\mbox{or}~~~M \hat{y} = b\:
\en
where we have multiplied each side of equation~(\ref{start}) by
$(A^T A)^{-1}$, the pixel hit counter. From now on, we also assume
that the noise in the timestream is stationary and that its covariance
matrix is normalised so that $\mbox{diag}~~N^{-1}_{tt'} = I$. The previous
change of variable allows us to subtract the sky signal from the data: 
since we have chosen a
resolution high enough to neglect the pixelisation noise, we have
indeed $ A P d \simeq A_{\infty} x_{\infty} + A P n $ and,
consequently, $d - A P d \simeq n - A P n$. The map making consists in
two step: first 
compute a simple co-added map from the time line, and second, solve
equation~(\ref{eqchgt}) for the {\it stripes map} $\hat{y}$. The
final optimal map is obtained by adding these two maps.

It is worth mentioning that the stripes map is completely
independent of the sky signal, as soon as the pixelisation noise
can be ignored. It depends only on the scanning strategy through the
matrix $A$ and on the noise characteristics through the matrix $N$.
Even if in principle this change of variable is irrelevant since it
does not change the matrix to be inverted, it does in practice
since $d-A P d$ is free from the fast variations of $d$ (up to the
pixelisation noise), as \eg the Galaxy crossings, which are
numerically damaging to the quality of the inversion.

To solve for equation \ref{eqchgt}, we follow the ideas of \cite{PrNe00}
and apply an approximate Jacobi relaxation scheme.
The Jacobi method consists in the following relaxation step to
solve for our problem
\be
{\hat y}^{n+1} = R {\hat y}^n + D^{-1} b~~~\mbox{and}~~~{\hat y}^0
= 0
\en
where $D$ is the diagonal part of the matrix $M$ and $R$ is the residual
matrix $R = I - D^{-1}M$. Computing the diagonal elements of
$M=P N^{-1}A$ is rather prohibitive. The idea of Prunet et al.
(2000) is to approximate $D \simeq I$ by neglecting the
off-diagonal elements of $N^{-1}$. The residual matrix then simplifies
greatly and writes
\be
R = P(I - N^{-1})A
\en
The approximate Jacobi relaxation step is therefore defined as
\be
{\hat y}^{n+1} = R {\hat y}^n + b ~~~\mbox{and}~~~{\hat y}^0
=0
\en
One clearly sees that if this iterative scheme converges, it is
towards the full solution of equation \ref{eqchgt}. To perform these
successive steps, it is extremely fruitful to remember the assumed
stationarity of the noise. Indeed whereas this assumption implies a circulant
noise covariance matrix in real space, it translates in Fourier space in the
diagonality of the noise covariance matrix. This is naturally another
formulation of the convolution theorem, since a stationary matrix
acts on a vector as a convolution, and a diagonal matrix acts as a
simple vector product, thus a convolution in real space is
translated as a product in Fourier space. The point is that the
manipulation of the matrix $N^{-1}$ is considerably lighter and
will be henceforth performed in Fourier space. Applying the matrix
$R$ to a map reduces then to the following operations in order

\begin{enumerate}
\item ``observe'' the previous stripes map $\hat y^n$
\item Fourier transform the resulting data stream
\item apply the low-pass filter $W=I-N^{-1}$
\item inverse Fourier transform back to real space
\item co-add the resulting data stream into the map $\hat y^{n+1}$.
\end{enumerate}

Assuming that the normalised noise power spectrum can be
approximated by $ P(f)= 1 +
\left(f_0/f\right)^{\alpha}$, the low-pass filter associated to each
relaxation step is given by
\be
W(f)=\frac{f_0^{\alpha}}{f_0^{\alpha}+f^{\alpha}}
\en
Since both $A$ and $P$ are norm-preserving operators, the norm of
the increment $\Delta \hat y ^{n} = \hat y ^{n}- \hat y
^{n-1}$ between step $n$ and $n+1$ decreases as  $\|\Delta \hat
y ^{n+1}\| \le W(f_{min}) \|\Delta \hat y ^{n}\|$, where $f_{min}$
is a minimal frequency in the problem. Since $W(f_{min}) < 1$, we
see that the approximate Jacobi relaxation scheme will converge in
every case, which is good news. On the other hand, since $W(f_{min})
\simeq 1$, the actual convergence rate of the scheme is likely to
be very, very slow, which is bad news (cf. figure~\ref{jacobi-residu} for a
graphical illustration) . The fact that this algorithm
is robust, but dramatically slow is a well-known property of the
Jacobi method. The multi-grid method is also well known  to solve this
convergence speed  problem. Note that if the convergence is reached, the solution we
get is the optimal solution, \ie similar to the one that would be
obtained by a full matrix inversion.

\subsubsection{Multigrid relaxation}
\label{mgaspect}

The multi-grid method for solving linear problems is described in
great details in 
\cite{PrTe92}. At each relaxation step at level $k=k_{max}$,
our target resolution, we can define the error $e_{k}^n=\hat
y_{k}^n-\hat y_{k}$ and the residual $r_{k}^n
= M_{k}\hat y_{k}^n-b_{k}$. Both are related through
\be \label{eqres}
M_{k} e_{k}^n = r_{k}^n
\en
If we are able to solve exactly for equation (\ref{eqres}), the
overall problem is solved. The idea of the multi-grid algorithm is
to solve approximately for equation (\ref{eqres}) using a coarser
``grid'' at resolution $k-1$, where the relaxation scheme should
converge faster. We thus need to define a fine-to-coarse operator, in
order to define the new problem on the coarse grid and solve for
it. We also need a coarse-to-fine operator in order to inject the
solution onto the fine grid. The approximation to the error
$e_{k}^n$ is finally added to the solution. The coarse grid
solver is usually applied after a few iterations on the fine level
have been performed (in practice we perform 3 to 5 iterations).
Naturally, since the solution to the problem on the coarse level
relies also on the same relaxation scheme, it can be itself
accelerated using an even coarser grid. This naturally leads to a
recursive approach of the problem.

We defined our fine-to-coarse operator to be an averaging of the values of
the 4 fine pixels contained in each coarse pixel. The
coarse-to-fine operator is just a straight injection of the value
of the coarse pixel into the 4 fine pixels. The most important issue
is the temporal rebinning of the data stream since the speed of the
iterative scheme at a given level is actually set by the two Fourier
transforms. We performed that resampling by simply taking each time we go
up one level one point out of two. At the lower resolutions, this
reduction is such that the iteration cost is negligible when
compared to that at higher $k$; it allows fast enough iterations
that full convergence may be reached. In practice we choose a minimal level
$k=3$ defined by $N_{side} = 8$ and iterate a few hundred times to
reach exact convergence.

Finally the navigation through levels allows several options to be
taken. Either we go up and down through all the levels successively
(the so-called ``V-cycle'') or we follow more intricate paths (\eg
the ``W-cycle'' where at a given level we go down and up all the
lower levels before going up and vice-versa). Since it turns out
that levels are relatively disconnected in the sense that the scales
well solved at a given resolution are only slightly affected by the
solution at a higher level, the ``V-cycle'' is appropriate and is
the solution we adopt in the two following configurations.

\section{Practical application to ARCHEOPS and TopHat experiments}
\label{application}

We now aim at demonstrating the capabilities of this algorithm with
simulated data whose characteristics are inspired by those of the ARCHEOPS and
TopHat experiments. 

\subsection{Simulation}

The ARCHEOPS and TopHat experiments are very similar with respect to
their scanning strategy. Indeed both use a telescope that simply spins
at a constant rate (respectively $3$ and $4$ RPM) about its vertical
axis. Thus due to Earth rotation the sky coverage of both is
performed through scan circles whose axis slowly drifts on the
sky. Nevertheless, because of the different launch points
(respectively on the Arctic circle (Kiruna, Sweden) and in Antarctica
(McMurdo)) and their different telescope axis elevation (respectively
$\sim 41^o$ or $12^o$) they do not have the same sky coverage.

Otherwise the two experiments do not use the same bolometers
technology, neither the same bands nor have the same number of
bolometers. But even if we try to be fairly realistic, our goal though
is {\em not} to compare their respective performances but rather to
demonstrate two applications of our technique in different
settings. We then simulate for each a data stream of $\sim 24\
\mathrm{hrs}$ duration with respectively a sampling frequency of $171$
and $64\ \mathrm{Hz}$. The TODs contain realistic CMB and Galactic
signal for a band at $143\ \mathrm{GHz}$. Note that this is a one day
data stream for TopHat (out of 10 expected) and that this frequency
choice is more appropriate for ARCHEOPS than for TopHat (whose
equivalent band is around $156\ \mathrm{GHz}$), but this is realistic
enough for our purpose. We generated a Gaussian noise time stream with
the following power spectrum $P(f) \propto
(1+({f_{knee}/f})^{\alpha})^{-1}$ with $f_{knee} = 0.24$ and $1\
\mathrm{Hz}$, and $\alpha = 1.68$ and $1$. The noise amplitude per
channel is choosen so that it corresponds for ARCHEOPS (24 hours) and
TopHat\footnote{Lloyd Knox private communication} (10 days of flight)
to $30 / 8\ \mu \textrm{K}$ on average per 
$20'$ pixel, with a beam FWHM of $10'$ / $20'$.

We introduced $5$ distinct levels of resolution defined by their
$N_{side}$ parameter in the HEALPix package \cite{GoHi98}. The
higher resolution level is imposed by the pixelisation noise level
requirement (section \ref{mgaspect}) to $N_{side} = 256$ (pixel size
$\sim 13.7'$) whereas the lower one is $N_{side} = 8$ (pixel size
$\sim 7.3^o$). Therefore these two configurations each offer an interesting
test since they differ by the sky coverage and the noise power
spectrum. We iterate $3$ times at each level except at the lowest one
where we iterate $100$ times.

\subsection{Results}

The algorithm is as efficient in both situations. Whereas for ARCHEOPS
whose timeline is longer due to the higher sampling frequency it took
2.25 hours on a SGI ORIGIN 2000 single processor, it took around 1.5
hours for the TopHat daily data stream.  

In figures \ref{rec_kir} and \ref{rec_top} we depict from top to
bottom and from left to right the initial co-added data, the
reconstructed noise map, the hit map, \ie the number of hit per pixels
at the highest worked out resolution,  the initial signal map as well as the
reconstructed signal map and the error map. We see that the destriping
is excellent in both situations and the signal maps recovered only
contain an apparently isotropic noise. We note the natural correlation
between the error map and the hit map.  Finally, we stress that
no previous filtering at all has been applied.

Figure \ref{level_compil} shows how the noise map is
reconstructed at various scales. This is a mere illustration of our
multi-grid work.

\begin{figure*} \begin{center} 
\end{center}
\caption{Simulated ARCHEOPS Kiruna flight : From top to bottom and
from left to right, the co-added map and the input Galactic + CMB
signal map, the reconstructed noise (stripes) and signal maps, the hit
count and the error map (Arbitrary unit). The fact that the coverage
is not fully symmetrical is due to the fact that we considered
slightly less than $24\mathrm{hr}$. Mollweide projection with pixels
of $13.7'$ (HEALPix $N_{side} = 256$). Arbitrary units.\label{rec_kir}}
\end{figure*}

\begin{figure*} 
\begin{center} 
\end{center}
\caption{ Simulated TopHat one day flight : From top to bottom and from
left to right, the co-added map and the input CMB + Galaxy signal map,
the reconstructed noise (or stripes) and signal map, the hit count and
error map. The fact that the coverage is not fully symmetrical is due
to the fact that we consider only $18.2 \mathrm{hr}$ of
flight. Gnomonic projection with pixel of $13.7'$ (HEALPix $N_{side} =
256$). Note that the slight visible stripping is correlated to the
incomplete rotation pattern. Arbitrary units.\label{rec_top}}
\end{figure*}

\begin{figure*}
\begin{center} 
\end{center}
\caption{Multi-grid noise map recovery : In this plot we show in the
ARCHEOPS Kiruna case how the noise map is reconstructed at various
levels, corresponding respectively to $N_{side} = 8,16,32,64$.\label{level_compil}}
\end{figure*}

\subsection{Tests}

At this point, some tests are required to cautiously validate our
method. First, as was stated below, as soon as the iterative algorithm
has converged, the solution is by construction the optimal solution,
similar to the one that would be obtained by the full matrix
inversion. As a criterium for convergence we required the $2$-norm of
residuals to be of the order of the machine accuracy.

\begin{figure}[!hbt] 
\psfig{figure=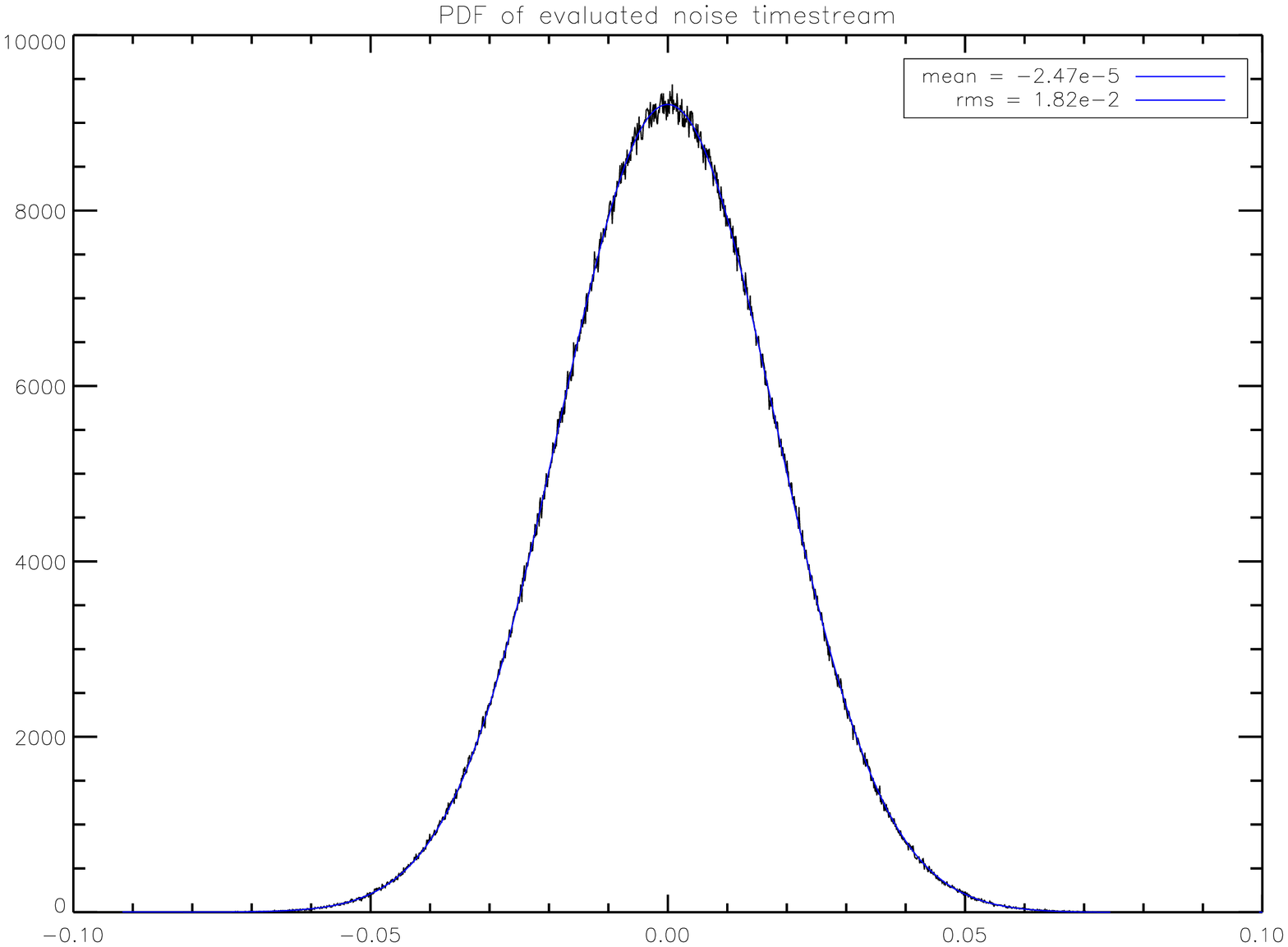,angle=90, width=\hsize}
\psfig{figure=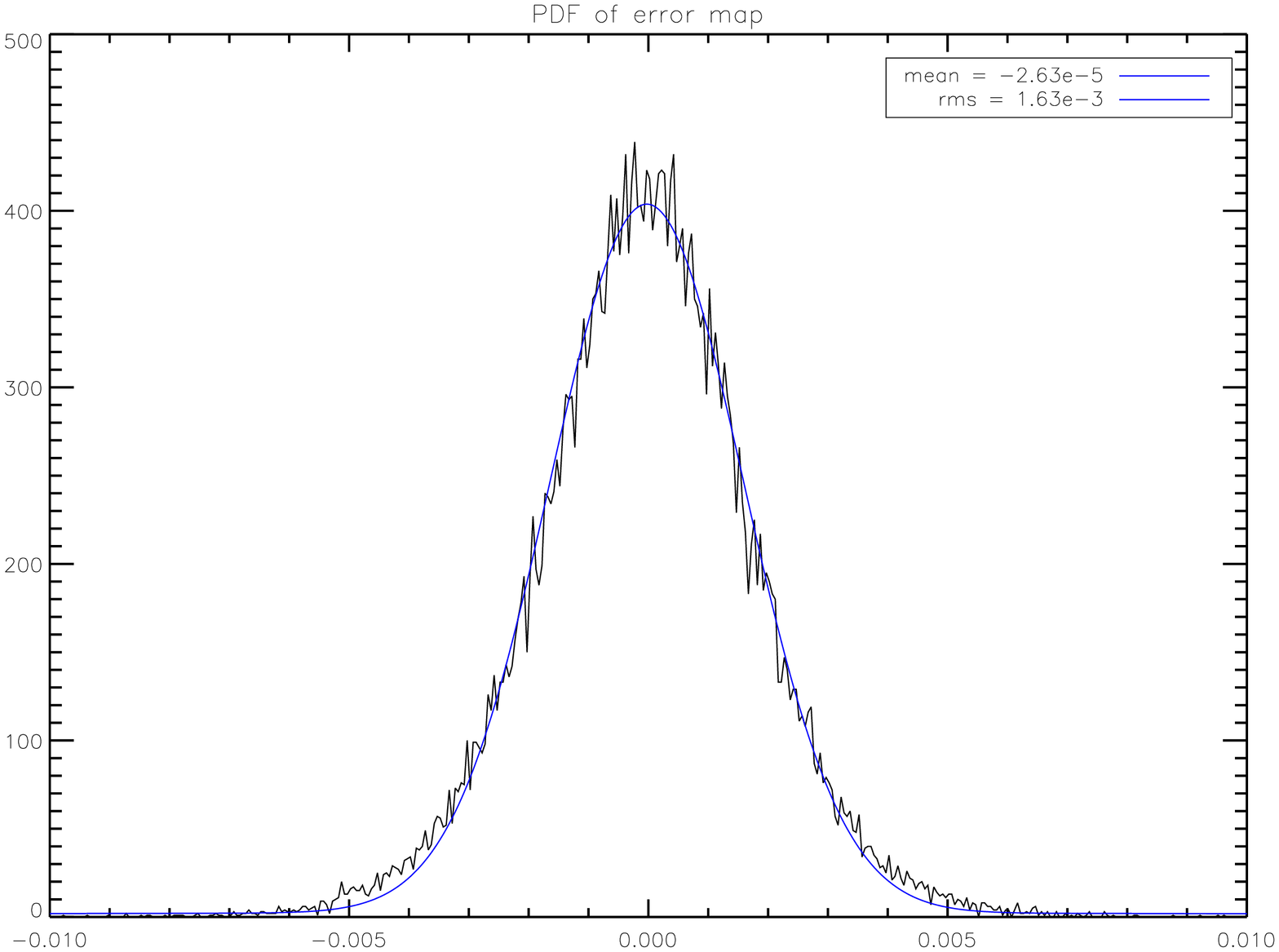,angle=90, width=\hsize}
\caption{ In the TopHat case we plot from top to bottom the recovered
probability distribution function of the noise stream evaluated along
the timeline as well as the error map PDF. In this two cases a fit to
a gaussian has been performed whose parameters are written inside the
figures. No significant departure from gaussianity are detected. The
arbitrary units are the same as the ones used for the previously shown
maps.\label{PDF}}
\end{figure}

\begin{figure}[!hbt] 
\psfig{figure=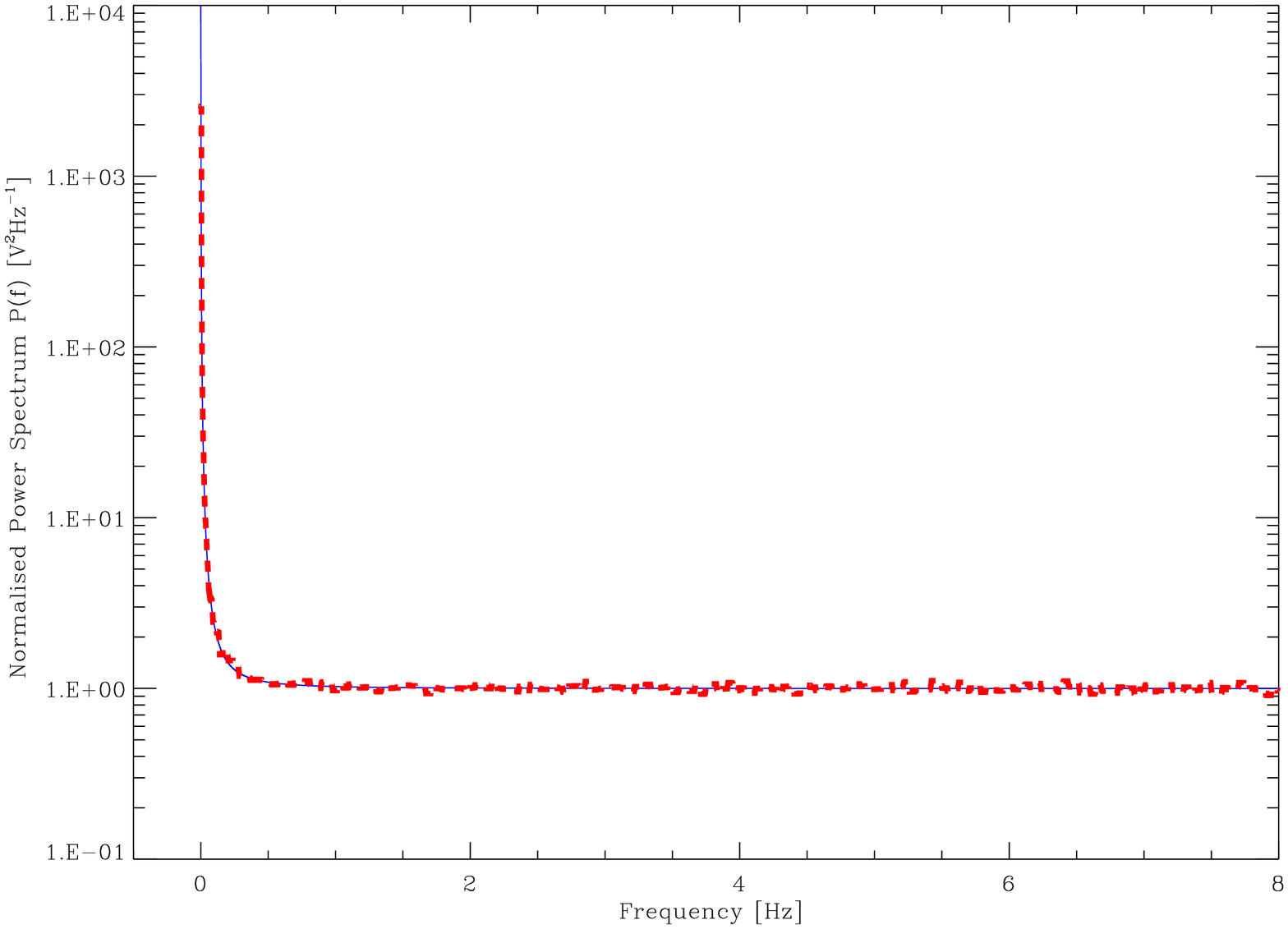,angle=90,width=\hsize}
\psfig{figure=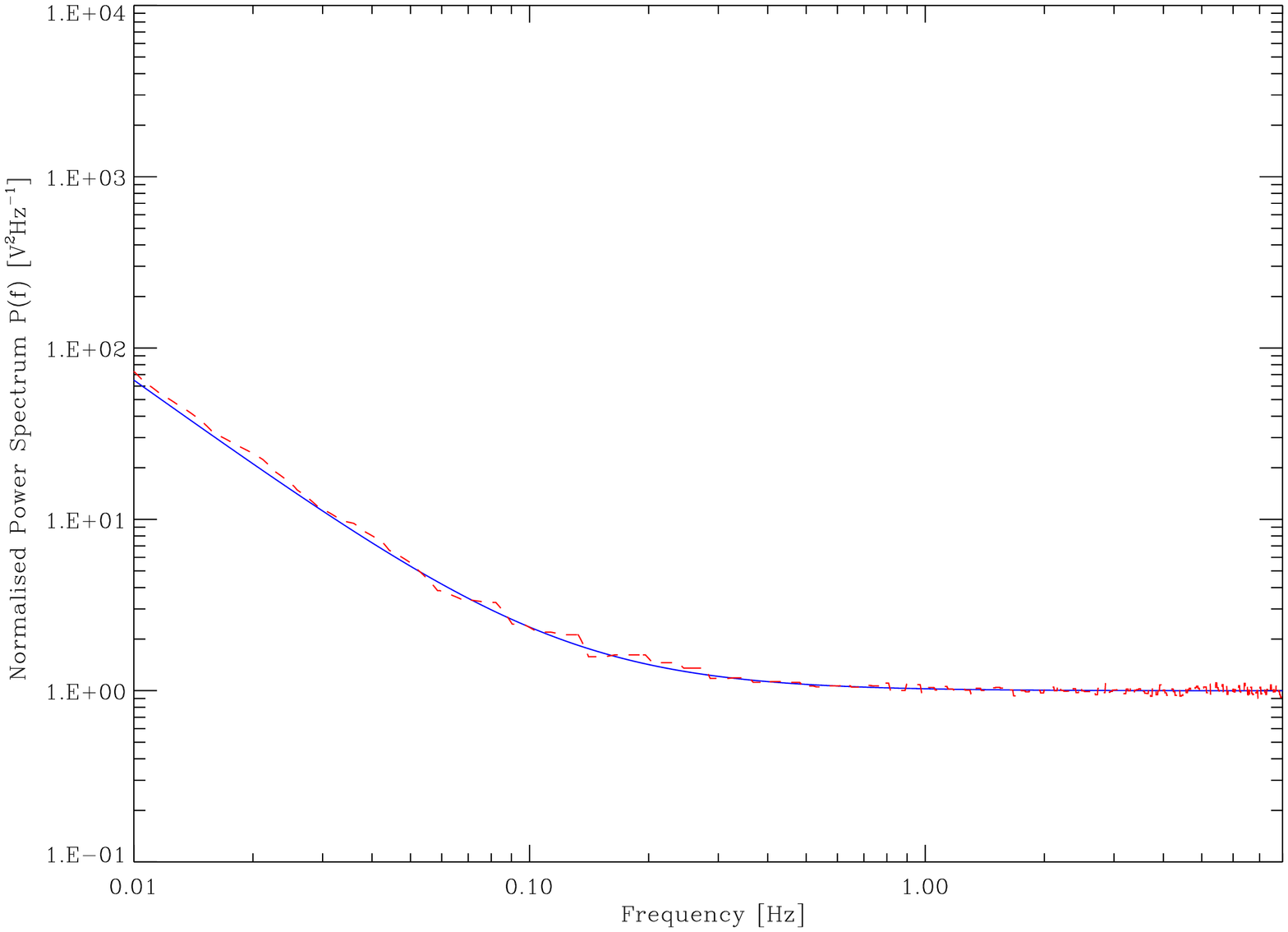,angle=90,width=\hsize}
\caption{Recovery of the noise power spectrum in the Archeops case
(top: linear x-axis, bottom: log x-axis): The red thin dashed line
shows the initial analytic noise power spectrum used to generate the
input noise stream and the blue thick line denotes the recovered one
after 6 iterations. The recovered one has been binned as described in
section \ref{noise_eval} and both have been normalised so that the
white high frequency noise spectrum is equal to one. The agreement is
obviously very good. No apparent bias is visible. Note that a perfect
noise power spectrum prior knowledge has been used in this
application.\label{spec_rec}}
\end{figure}

We initially have a Gaussian random noise stream fully characterised
by its power spectrum. Therefore an important test is to check wether
the deduced noise stream (by ``observing'' the stripe map, see
\S~\ref{noise_eval} for further details definition) is Gaussian and
has the same power spectrum. On figure \ref{PDF} we ensure that the
evaluated noise time stream is indeed gaussian.  As depicted on figure
\ref{spec_rec}, where we plot in the ARCHEOPS case both the analytical
expression of the spectrum according to which we generate the timeline
and its recovered form, the agreement is excellent. We recall that we
assumed at the beginning a perfect knowledge of the noise in order to
define the filters. This is naturally unrealistic but the issue of
noise evaluation is discussed in section \ref{noise_eval} below. We
plotted as well the probability distribution function (PDF) of the
final error map, \ie the recovered noisy signal map minus the input
signal map (figure \ref{PDF}). This PDF is well fitted by a gaussian whose
parameters are given in figure \ref{PDF}. The PDF of the error map
displays some non-gaussian wings. Let us recall here that this is no
surprise here because of the non-uniform sky coverage as well as the
slight residual stripping, both due to the non-ideal scanning
strategy, \ie that produces a non-uniform white noise in pixel space.

Another particularly important test consists in checking the
absence of correlation between the recovered noise map and the
initial signal map. We could not find any which is no surprise since
we are iterating on a noise map (see \S~\ref{noiseiter}) which does not contain any signal up
to the pixelisation noise, that is ensured to be negligible with regards to the
instrumental noise by choice of the resolution $k _{max}$ (see \S~\ref{lmax}).

\section{Discussion}
\label{discussion}

\subsection{Why is such an algorithm so efficient ?}
\label{whygood}
The efficiency of such a method can be intuitively understood. Indeed,
although the Jacobi method is known to converge very safely it suffers
intrinsically from a very slow convergence for large-scale
correlations (which originate mainly in the off diagonal terms of the
matrix to be inverted) \cite{PrTe92}. This is illustrated on figure
\ref{jacobi-residu}: there we show the maps of residuals after solving this
system using a \emph{standard} Jacobi method on simulated data with
$50$, $100$, $150$, and $200$ iterations. We used the same simulation
and therefore the same sky coverage. Obviously the largest structures
are the slowest to converge (implying observed large scale residual
patterns). As a consequence it seems very natural to help the
convergence by solving the problem at larger scales. Whereas
large-scale structures will not be affected by a scale change, smaller
structures will converge faster.

\begin{figure*}
\begin{center} 
\end{center}
\caption{Residual map after 50, 100, 150 and 200 iterations of a
standard Jacobi method. This has been performed on simulated data
for one bolometer with a nominal noise level. The sky coverage is
that of ARCHEOPS coming Kiruna flight. The residual large scale
patterns illustrate the difficulties the standard Jacobi method faces
to converge. The stripes free area are the ones of scan crossing (see
the hit map in figure \ref{rec_kir}).\label{jacobi-residu}}
\end{figure*}

\subsection{Scalings}

\begin{figure}[!hbt]
\psfig{figure=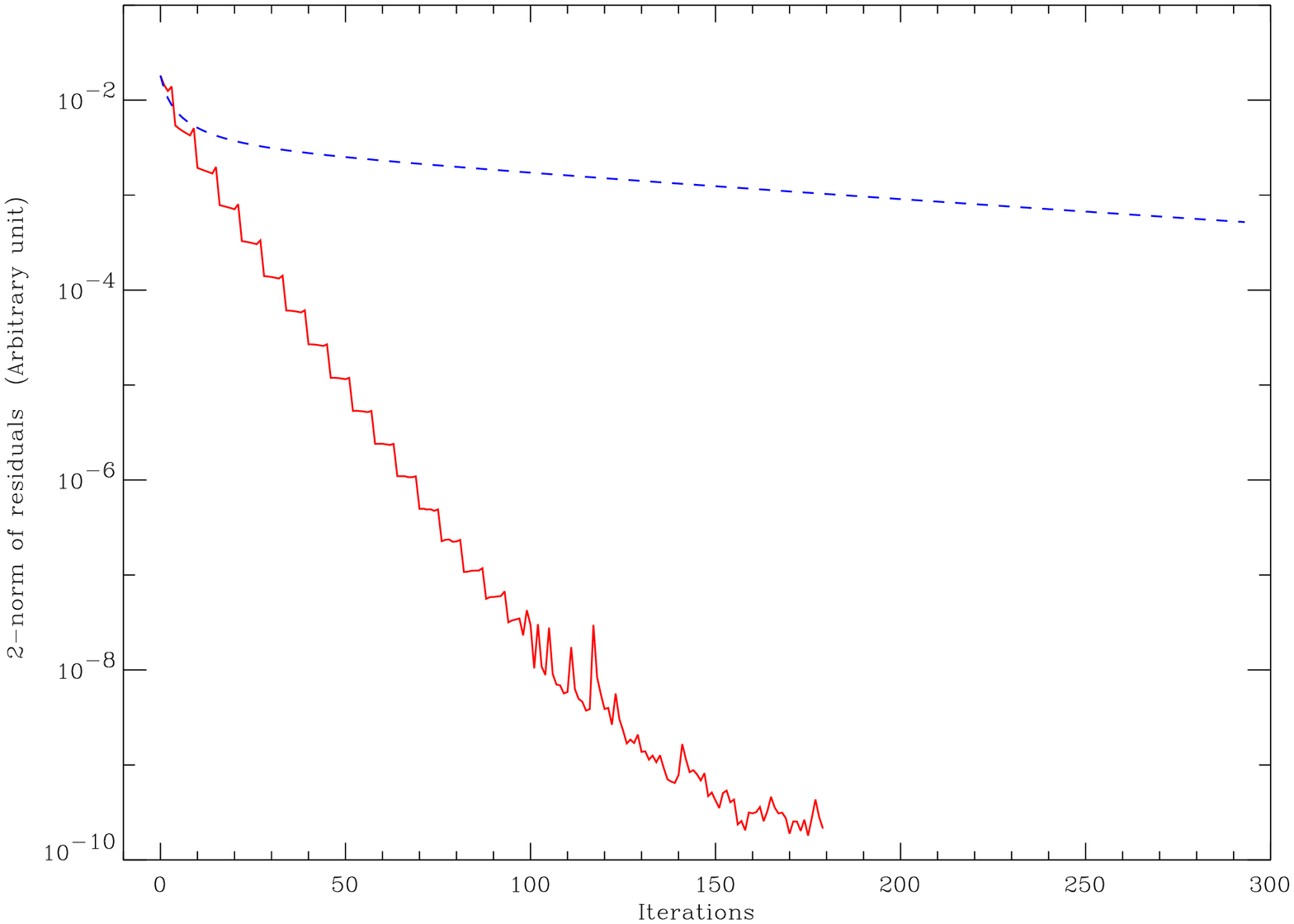, angle=90, width=\hsize}
\caption{The evolution of the 2-norm of residuals with the number of
iterations at level max. Whereas the blue dashed line is standard
iterative Jacobi, the solid red line is the iterative multi-grid
Jacobi method. A full multi-grid cycle incorporates 3 iterations at
level max before going down and up all the levels. The sharp jumps
corresponds to the moment when the scheme reach again level max and
thus take the full benefits from the resolution at lower levels. Note
that the very sharp features after $\sim 100$ iterations are due to
the fact we reached the machine accuracy which is almost out of reach for a
standard iterative method.\label{conv-rate}}
\end{figure}

We have found that this multi-grid algorithm translates naturally in
\emph{a speed up greater than $10$} as compared to a standard Jacobi
method. This is illustrated in figure (\ref{conv-rate}) where we
plotted the evolution of the 2-norm of residuals for the two methods
in terms of the number of iterations in 'cycle units'. One cycle
corresponds to 8 iterations at level $k_{max}$ for a standard
Jacobi method whereas it incorporates addionally going up and down all
the lowest levels in the multi-grid case. Thus the cycle timing is not
exactly the same but the difference is negligible since the limiting
stages are definitely the iterations performed at maximum
resolution. Note the fact that the efficiency of the multi-grid method
allows us to solve exactly the system up to the machine accuracy
(small rebounds at the bottom of the curve) in approximately
$135\textrm{mn}$ for $(\mathcal{N}_{tod} \sim 8\ 10^6,
\mathcal{N}_{pix} \sim 8\ 10^5)$ on a SGI ORIGIN 2000 single
processor. Since the limiting stages are the FFT's at the higher
levels, this algorithm scales as $\mathcal{O} (\mathcal{N}_{tod} \ln
\mathcal{N}_{tod})$. In terms of memory storage it scales naturally as
$\mathcal{O} (\mathcal{N}_{tod})$ since one crucial feature of this
iterative method is to handle only vectors and never an explicit
matrix. The scaling of this problem is formally independent of the
number of pixels $\mathcal{N}_{pix}$.

\subsection{The noise estimation issue and noise covariance matrix estimation}
\label{noise_eval}

\begin{figure}[!bt]
\psfig{figure = 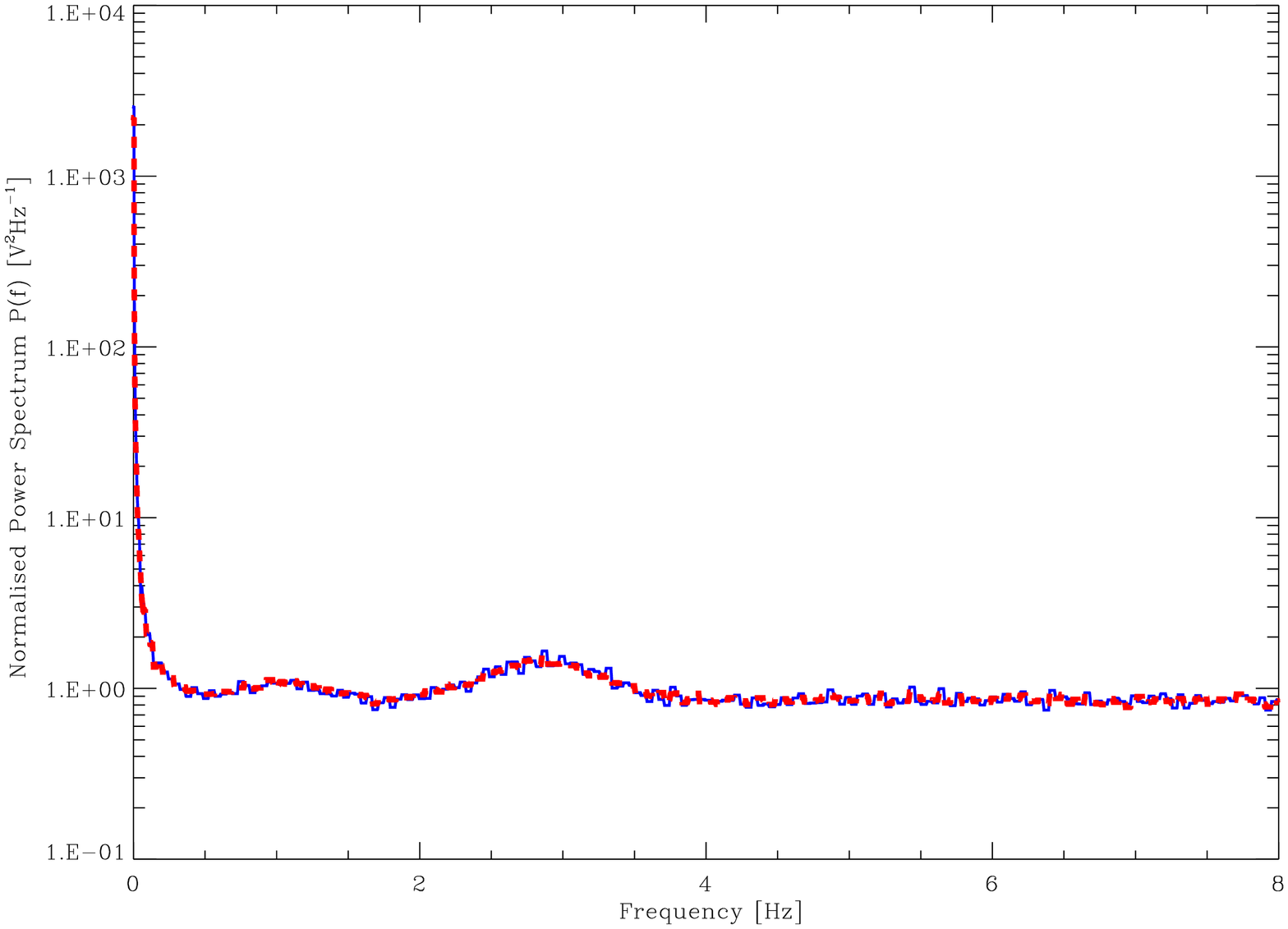,angle=90,width=\hsize}
\psfig{figure = 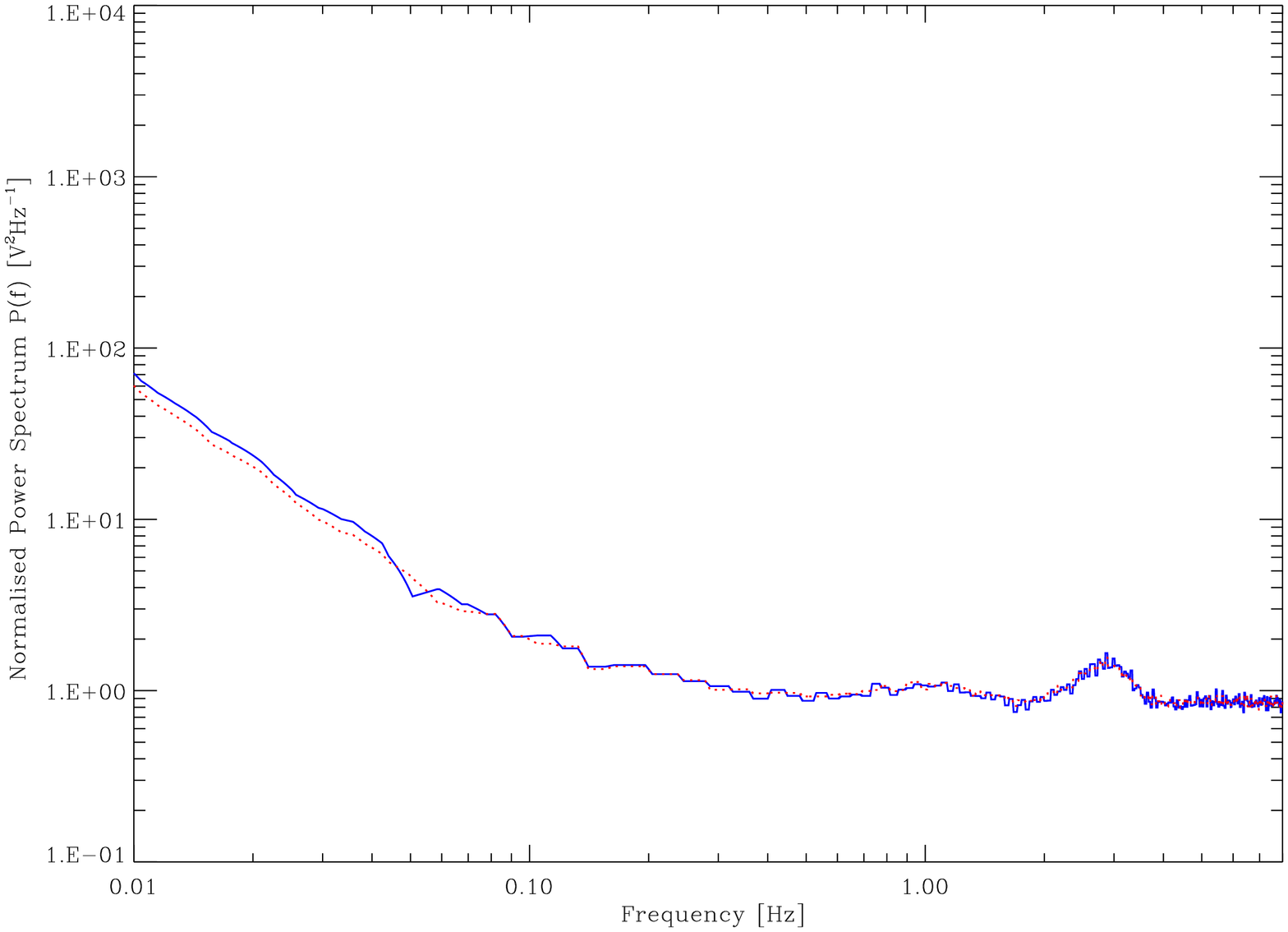,angle=90,width=\hsize}
\caption{Evaluation of the noise power spectrum (top: linear x-axis,
bottom: log x-axis): the red thin dashed line shows the initial noise
power spectrum obtained from the input noise stream and the blue thick
line denotes the recovered one after 5 iterations. Both have been
smoothed and normalised so that the white high frequency noise
spectrum is equal to one. The agreement is obviously very good. Note
that no noise priors at all have been used in this evaluation.\label{noise_eval_1}}
\end{figure}

The estimation of the statistical properties of the noise in the
timestream is an important issue for this kind of algorithm. Indeed,
whereas till now we have assumed a perfect prior knowledge of the
noise power spectrum in order to define the filters, it might not be
that easy in practice since we can not separate the signal from the
noise. We will therefore aim at making a joint estimation of the
signal and the noise. This has been pioneered recently by
\cite{FeJa00} and implemented independently by \cite{PrNe00}. The
latter implementation is rather straightforward in our particular case
since it just implies reevaluating the filters after a number of
iterations, given the (current) estimation of the signal map and thus
of the signal data stream. Nevertheless its non-obvious convergence
properties have to be studied carefully through simulations. Making
use of (\ref{noisedef}) our evaluation of the noise timeline
$\hat{n}^n$ at the $n^{th}$ iteration and at level max is \be
\hat{n}^n = d -A(\hat{y}^{\ n}_{k _{max}} + P d) \: .  \en Then we
compute its spectrum and (re-) define the required filters. We then go
through several multi-grid cycles (5 in the above demonstrated case)
before re-evaluating the noise stream. Very few evaluations of the
noise are needed before getting a converged power spectrum (around
2). In such an implementation, no noise priors at all are
assumed. This is illustrated on one particular worked out example in
the case of a 4 hours ARCHEOPS like flight (more detailed
considerations will be discussed somewhere else). To reduce the number
of degrees of freedom we bin the evaluated noise power spectrum using
a constant logarithmic binning ($\Delta \ln f=0.15$ in our case) for
$f \leq 2~f_{knee}$ and a constant linear binning
($\Delta~f~=~0.08~\mathrm{Hz}$ in our case) for higher frequency. The
figure \ref{noise_eval_1} shows the genuine and evaluated noise power
spectrum. The initial noise power spectrum was a realistic one
$P(f)~\propto~(1+({f_{knee}/f})^{\alpha}) \displaystyle$ to which we
added some small perturbations (the two visible bumps) to test the
method. Note the small bias around the telescope spin frequency at
$f_{spin} = 0.05 \mathrm{Hz}$: this is illustrative of the
difficulties we fundamentally face to separate signal and noise at
this particular frequency through equation~(\ref{noisedef}) . Naturally, this
bias was not present in the case demonstrated on figure \ref{spec_rec}
where we assumed a prior knowledge of the spectrum. This possible bias
forced us to work with a coarser binning ($\Delta \ln f=1.$) in the
$1/f$ part of the spectrum till the convergence is reached, \ie we
evaluate the noise power spectrum with the previously mentioned
binning only at the last step. Proceeding this way, the convergence
towards the correct spectrum is both fast ($3$ noise evaluations) and
stable.

\begin{figure}[!hbt]
\psfig{figure=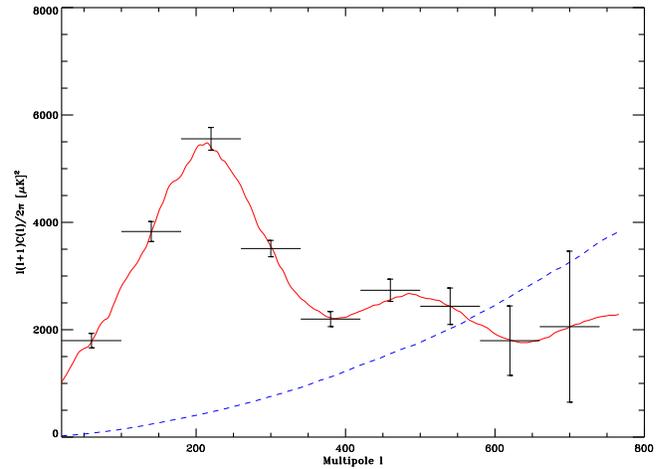, angle=90,width=\hsize}
\caption{In the case of a \emph{one day} flight of the TopHat
experiment we perform a very approximate evaluation of the recovered
signal band powers (black) which has to be compared to the input
signal power spectrum (red line). Both have been measured the same way
using the \texttt{anafast} routine of the HEALPix package. These band
powers $(\Delta \ell = 80)$ have been performed using a fast
Monte-Carlo evaluation of $C_{\ell}^{noise}$ (dashed blue line) which
does not correspond exactly to white noise since there remains some
spatial correlations. This constitutes in no way an appropriate
$C_{\ell}$ measurement but is an encouraging step towards a full
Monte-Carlo treatment.\label{cl_mc}}
\end{figure}

Second, the output of any map-making should contain as well an
evaluation of the map noise covariance matrix $(A^T N^{-1}
A)^{-1}$. Given such a fast algorithm and given an evaluation of the
power spectrum, it is natural to obtain it through a Monte-Carlo
algorithm fueled with various realizations of the noise timeline
generated using the evaluated power spectrum. This part will be
presented in a future work. However we illustrate it very briefly by a
very rough $C_{\ell}$ determination (which is in no way an appropriate
$C_{\ell}$ estimate). To this purpose we perform a one day TopHat like
simulation including only the CMB signal plus the noise. From this
data stream we obtain an ``optimal'' signal map as well as an
evaluation of the noise power spectrum using the previously
described algorithm. With the help of the \texttt{anafast} routine of
the HEALPix package we calculate this way a rough
$C_{\ell}^{signal}$. Using the estimated noise power spectrum we
generate $10$ realisations of the noise and get consequently $10$
``optimal'' noise maps. For each of them we measure as before
$C_{\ell}$ and average them to obtain $C_{\ell}^{noise}$. In order to debiase the signal power
spectrum recovered in this way, we substract $C_{\ell}^{noise}$ to
$C_{\ell}^{signal}$. The power spectrum obtained in this manner
includes as well some spatial correlations due to some light residual
stripping and thus does not correspond to white noise (at least in the
low $\ell$ part). For the aim of comparison we compute the power
spectrum of the input signal measured the same way, $C_{\ell}^{input}$,
and plotted both $C_{\ell}^{input}$ as well as $C_{\ell}^{signal} -
C_{\ell}^{noise}$ averaged in linear bands of constant width $\Delta
\ell = 80$. The agreement is obviously very good as illustrated on
figure \ref{cl_mc}. The error bars take into account both the sampling
induced variance as well as the beam spreading \cite{Kn95}. A few
comments need to be made at this point. This is 
in no way an appropriate $C_{\ell}$ measurement since we are not
dealing properly here with the sky coverage induced window function
which triggers some spurious correlations within the $C_{\ell}$s. 
We thus do not take into account the full-structure of the map
noise covariance matrix. Nevertheless, the efficiency of such a rough
method is encouraging for more detailed implementation and a full
handling of the noise covariance matrix.
 
Note finally that this is a naturally parallelised task which should
therefore be feasible very quickly.

\subsection{Application to genuine data and hypothesis}

The application to genuine data could be problematic if our key
hypothesis were not to be fulfilled thus we have to discuss them.
Concerning the noise, we assumed that it is Gaussian in order to derive
our map estimator and stationary in order to exploit the diagonality
of its noise covariance matrix in Fourier space. Both hypothesis are reasonable
for a nominal instrumental noise, at least partially on (sufficiently
long) pieces of timeline. If not, we would have to cut the bad parts and
replace them by a constrained realization of the noise in these
parts. Concerning the signal, no particular assumptions are needed in
the method we are presenting. At this level, we neglected as well the
effect of the non perfect symmetry of the instrumental beam. This
effect should be quantified for a real experiment
\cite{Wu00,Do00}. Another technical hypothesis is the negligibility of
the pixelisation noise with respect to the instrumental noise but
since this is fully under control of the user it should not be a
problem.

In this paper, we have presented an original fast map-making method
based on a multi-grid Jacobi algorithm. It naturally entails the
possibility of a joint noise/signal estimation. We propose a way to
generate the noise covariance matrix and illustrate its ability on a
simple $C_{\ell}$ estimation. The efficiency of this
method has been demonstrated in the case of two coming balloon
experiments, namely ARCHEOPS and TopHat but it naturally has a more
general range of application.  This tool should be of natural interest
for Planck and this will be demonstrated somewhere else. Furthermore,
due to the analogy of the formalisms, this should have some
applications as well in the component separation problem.

We hope to apply it very soon on ARCHEOPS data. The \texttt{FORTRAN
90} code whose results have been presented is available at {\bf
\texttt{http://ulysse.iap.fr/download/mapcumba/}}. 

\acknowledgements

We would like to acknowledge Eric Hivon for fruitful remarks, Lloyd
Knox for stimulating discussions and information on the
TopHat experiment and the developers of the HEALPix \cite{GoHi98}
package which has considerably eased this work.


\end{document}